\documentclass[twocolumn,prl]{revtex4}
\usepackage{amssymb}
\usepackage{graphicx}
\usepackage{dcolumn}
\usepackage{bm}
\usepackage{amsmath}

\begin{document}

\title{Decoherence and dephasing in strongly driven colliding Bose-Einstein condensates}
\author{N. Katz, R. Ozeri, E. Rowen, E. Gershnabel and N. Davidson}
\affiliation{Department of Physics of Complex Systems,\\
Weizmann Institute of Science, Rehovot 76100, Israel}

\begin{abstract}

We report on a series of measurements of decoherence and
wavepacket dephasing between two colliding, strongly coupled,
identical Bose-Einstein condensates. We measure, in the strong
excitation regime, a suppression of the mean-field shift, compared
to the shift which is observed for a weak excitation. This
suppression is explained by applying the Gross-Pitaevskii energy
functional. By selectively counting only the non-decohered
fraction in a time of flight image we observe oscillations for
which both inhomogeneous and Doppler broadening are suppressed, in
quantitative agreement with a full Gross-Pitaevskii equation
simulation. If no post selection is used, the decoherence rate due
to collisions can be extracted, and is in agreement with the local
density average calculated rate.
\end{abstract}

\maketitle

The dephasing of a momentum excitation in a non-uniform
Bose-Einstein condensate (BEC) is governed by several factors
\cite{structure-factor}, \cite{ketterle-bragg}. Notably, the
inhomogeneous broadening of the Bogoliubov energy shift and the
Doppler broadening. We denote these processes as dephasing since
they involve a reversible, although nonlinear, evolution of the
macroscopic wavefunction. In contrast, the decoherence of such an
excitation due to collisions with the BEC, involves coupling to a
quasi-continuum of initially unoccupied momentum states
\cite{ours-beliaev}. This process is driven by the fluctuating
ground-state occupation of these modes, and is therefore
inherently irreversible on timescales larger than the so called
"memory time" of the system \cite{cohen}.

\begin{figure}[tb]
\begin{center}
\includegraphics[width=8cm]{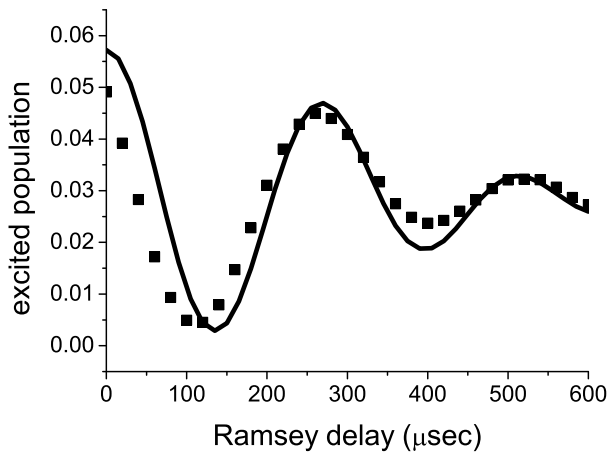}
\end{center}
\caption{Ramsey dephasing calculation. We simulate, by solving the
cylindrically symmetric GPE \cite{GPE}, a Ramsey type experiment
(solid boxes). We apply a detuned ($\sim 3$ kHz) and weak $40$
$\mu$sec Bragg pulse, wait a variable Ramsey delay time, and then
apply a second pulse which is in-phase with the first. A
simplified local density averaging is also shown (solid line,
described in the text), and is seen to capture the essence of the
more complete simulation. Both curves decay at roughly the
expected \cite{structure-factor} inhomogeneous dephasing rate
($3.0$ kHz). \cite{lineshapes} } \label{fig1}
\end{figure}

In previous works the inhomogeneous lineshape was seen to agree,
for short Bragg coupling times, with both the experimental data
\cite{ketterle-phonon}, \cite{ours-multibranch}, and with
simulations of the Gross-Pitaevskii equation (GPE)
\cite{dalfovo-momentum}. At longer timescales the inhomogeneous
lineshape is resolved into radial modes \cite{ours-multibranch}.
Weak probe Bragg spectroscopy and interferometry were also used to
characterize the coherence and spatial correlation function of
condensates \cite{phillips} and quasi-condensates
\cite{linewidths}. In this letter we measure the dephasing and
decoherence of strongly driven oscillations between two identical
colliding BECs. By post selecting the non-decohered fraction we
observe a strong suppression of both inhomogeneous and Doppler
broadening mechanisms, which is in quantitative agreement with the
results of a GPE simulation. The observed collisional decoherence
rate is seen to agree with the expected local density
approximation (LDA) average of the free particle collision rate.

The RMS width of the LDA inhomogeneous lineshape for high momentum
excitations is given by $\sqrt{8/147}\mu$, where $\mu$ is the
chemical potential of the BEC \cite{structure-factor}. The RMS
Doppler broadening is given by, $\sqrt{8/3}\hbar k/mR_{z}$
\cite{structure-factor}, where $k$ is the wavenumber of the
excitation, $m$ is the mass of the BEC atom and $R_{z}$ is the
Thomas-Fermi radius of the condensate. For our experimental
parameters the Doppler dephasing rate is predicted to be $0.6$
kHz, which is clearly dominated by the expected $3.0$ kHz
inhomogeneous dephasing rate \cite{lineshapes}. A GPE simulation
\cite{GPE} of the final excited population in a weak Bragg/Ramsey
type experiment (an initial $40$ $\mu$sec weak coupling Bragg
pulse, followed by a delay, and then a second pulse in phase with
the first), with simulation parameters as described for our
experimental system below, is shown in Fig. 1 (solid boxes). A
simplified LDA of the Ramsey signal (solid line), taking only the
density inhomogeneity into account, is seen to be in agreement
with the full simulation. Here the weakly excited state, in the
Bloch vector picture, precesses around the effective inhomogeneous
local detuning, and this leads to decaying oscillations in the
total final excited population. The calculated inhomogeneous
dephasing rate of $3.0$ kHz for our system agrees roughly with the
Ramsey signal both in the simulation and in LDA theory. We
conclude that weak excitations, at short times, are well
understood by LDA theory \cite{phillips}.

We now turn to the case of a strong excitation, i.e. where
population of the zero-momentum condensate is completely (or
nearly) depleted and transferred coherently to a travelling
condensate in a Rabi-like oscillation. For sufficiently large $k$,
the momentum of the excitation, we approximate such states by
$\psi(r,z)=a \phi_{0}(r,z)+ b e^{ikz}\phi_{0}(r,z)$
\cite{Meystre}, where $a$ and $b$ are the amplitudes of the
effective two level system and $|a|^2+|b|^2=N$, with $N$ being the
total number of atoms. $\phi_{0}(r,z)$ is the ground-state
wavefunction of the system with no excitations. The energy of such
a state $\psi$ can be evaluated by the Gross-Pitaevskii energy
functional \cite{CCT2} $E=\int dV \{\frac{\hbar^2}{2m}|\nabla
\psi(r,z)|^2+V_0(r,z)|\psi(r,z)|^2+\frac{g}{2}|\psi(r,z)|^4\}$,
where $V_0$ is the external potential and $g$ is the mean-field
coupling constant. By differentiating the energy by the excited
state population $|b|^2$, and considering the result as a function
of $|b|^2$, we find a population dependant excitation energy. For
our experimental system we find a nearly linear decrease in
excitation energy from the Bogoliubov prediction (weak excitation)
at low $|b|^2$, via the free particle value at $|b|^2=N/2$ down to
a symmetrically downshifted energy for $|b|^2 \approx N$, where
the zero-momentum state is simply a weak excitation of the
travelling condensate. The intuitive picture is that the energy
per particle required to transfer the condensate from $|b|^2=0$ to
$|b|^2=N$ is simply the free particle energy, since the internal
interaction energy is not changed by such a transformation. At
sufficiently large Rabi frequencies, that achieve complete
population inversion, the temporary mean field detuning averages
to zero due to its symmetric nature, and the resonance shifts to
the free particle value.

Our experimental apparatus is described in \cite{Steinhauer}.
Briefly, a nearly pure ($>90\%$) BEC of $1.6 (\pm 0.5)\times
10^{5}$ $^{87}$Rb atoms in the $|F,m_{f}\rangle =|2,2\rangle $
ground state, is formed in a QUIC\ type magnetic trap \cite{QUIC}.
The trap is cylindrically symmetric, with radial ($\hat{r}$) and
axial ($\hat{z}$) trapping frequencies of $2\pi \times 226$ Hz and
$2\pi \times 26.5$ Hz, respectively. This corresponds to
$\mu/h=2.48$ kHz.

\begin{figure}[tb]
\begin{center}
\includegraphics[width=8cm]{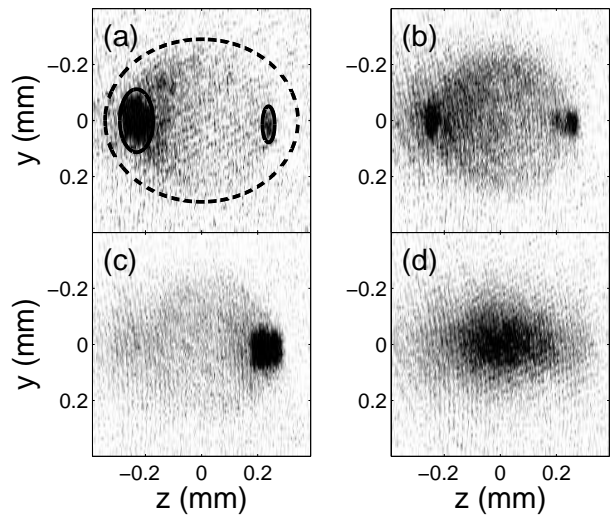}
\end{center}
\caption{Time of flight images of oscillating and colliding BECs.
(a) A weak perturbation of the BEC (left, in solid line ellipse)
to the excited population (right, in solid line ellipse). The
dashed line marks the region of interest for measuring the average
momentum $P_{tot}$ (b) A $\pi/2$ pulse, note the strong
collisional sphere. (c) Almost a $\pi$ pulse, note the weak
collisional sphere, indicating collisions between the excitation
and the zero momentum BEC. (d) After further oscillation ($>10\pi$
at $8.6$ kHz), the BECs are completely decohered, and the Bragg
coupling no longer effects the system.} \label{fig2}
\end{figure}

We excite the condensate at a well defined wavenumber using
two-photon Bragg transitions \cite{Phillips-Bragg}. The two Bragg
counter-propagating (along $\hat{z}$) beams are locked to a
Fabri-Perot cavity line, detuned 44 GHz below the
$5S_{1/2},F=2\longrightarrow 5P_{3/2},F^{\prime }=3$ transition.
At this detuning and at the intensities used here, there are no
discernable losses from the condensate due to spontaneous
emission. The frequency difference $\delta \omega$ between the two
lasers is controlled via two acousto-optical modulators. Bragg
pulses of variable duration and intensity are applied to the
condensate, controlling the excitation process.

Following the Bragg pulse, the magnetic trap is rapidly turned
off, and after a 38 msec of time of flight expansion the atomic
cloud is imaged by an on-resonance absorption beam, perpendicular
to the $\hat{z}$-axis. Fig. 2 shows the resulting absorption
images, for different excitation strengths and duration. Fig. 2a
shows a perturbative excitation with the large cloud at the left
corresponding to the BEC. A halo of scattered atoms is visible
between the BEC and the cloud of unscattered outcoupled
excitations to the right. In Fig. 2b we show a $\pi/2$ pulse which
generates a nearly symmetric excitation. The two condensates
collide producing a strong collisional sphere. In Fig. 2c we show
a nearly complete $\pi$ pulse, with a weak zero momentum component
remaining as an excitation of the travelling condensate. We note
the weak thermal cloud surrounding the origin, which is largely
unaffected by the Bragg pulse. When we increase the duration time
of Rabi oscillations (Fig. 2d), the effective two level system is
eventually depleted by collisions.

\begin{figure}[tb]
\begin{center}
\includegraphics[width=8cm]{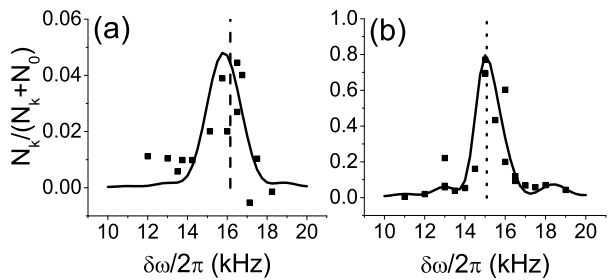}
\end{center}
\caption{Measured Bragg response (solid boxes) spectroscopy
($N=0.9\times 10^5$, and pulse duration of 0.5 msec for these
measurements). Solid lines are GPE simulations (a) Weak excitation
spectrum exhibiting the usual mean-field shifted Bogoliubov
excitation resonance. The dashed line is at the theoretical LDA
excitation energy ($16.15$ kHz). (b) Strong excitation spectrum,
in which a clear suppression of the mean-field shift is observed.
The dotted line is at the free particle excitation energy ($15.08$
kHz).} \label{fig3}
\end{figure}

We measure the response of our system by integrating over the
elliptical areas shown in Fig. 2a. The solid lines are gaussian
fit areas to the uncollided atoms. The dashed line contains the
entire region of interest, including the collisional products. We
define the number of atoms observed in the excited region inclosed
in the ellipse to the right as $N_{k}$ and those corresponding to
the initial BEC (left) as $N_{0}$. $P_{tot}$ is the total momentum
(in units of the momentum of a single excitation) along $\hat{z}$
measured inside the dashed ellipse, and normalized by the overall
number of observed atoms.

The suppression of the mean-field shift in strongly excited
condensates is shown in Fig. 3b, where the resonance is clearly
($14.9 \pm 0.2$ kHz from a gaussian fit) in agreement with the
free particle value ($15.08$ kHz, indicated by the dotted line).
This should be compared to the resonance ($16.05 \pm 0.2$ kHz from
a gaussian fit) observed in Fig. 3a for a weak excitation in
agreement with the expected LDA value of $16.15$ kHz (indicated by
the dashed line). The solid lines are GPE simulations of the
system, which also confirm the suppression of the mean-field
shift.

This suppression of the mean-field shift should cause a similar
decrease in the inhomogeneous broadening, leading to longer
coherence times for strongly driven condensates. We explore this
at various driving Rabi frequencies (but holding $\delta
\omega=2\pi \times 15$ kHz constant). Fig. 4a and 4b show
$P_{tot}(t)$ at the driving frequencies of $1.2$ kHz and $8.6$
kHz, respectively. Fig. 4c and 4d show $N_k/(N_0+N_k)(t)$, for the
same driving frequencies.

The dashed lines in Fig. 4a and 4b are exponentially decaying
oscillations, fitted to the experimental points. Here the decay is
mainly due to collisions between the two condensates. The solid
lines in Fig. 4c and 4d are GPE simulations with no fit
parameters. We observe the remarkable result that by
post-selecting the non-collided fraction we can recover the GPE
dephasing behavior, despite the fact that the GPE totally
disregards collisions between the excitations and the BEC and
considers only the evolution of the macroscopic wavefunction.

The results of this analysis are summarized in Fig. 5. Here the
solid boxes show the measured dephasing decay rate of
$N_k/(N_0+N_k)(t)$. An additional experimental point is measured
at the Bragg driving frequency of $3.4$ kHz. The solid line
represents the theoretical dephasing rate obtained by numerically
solving the GPE and then fitting the time evolution to an
oscillating exponential decay.

The empty boxes are the fitted decay rate of the oscillations in
$P_{tot}(t)$. The dashed line is the result of the theoretical sum
of the GPE dephasing rate calculated above (solid line), plus the
free particle LDA collision rate (1.66 kHz)
\cite{collision-value}, with no fitting parameters.

\begin{figure}[h]
\begin{center}
\includegraphics[width=8cm]{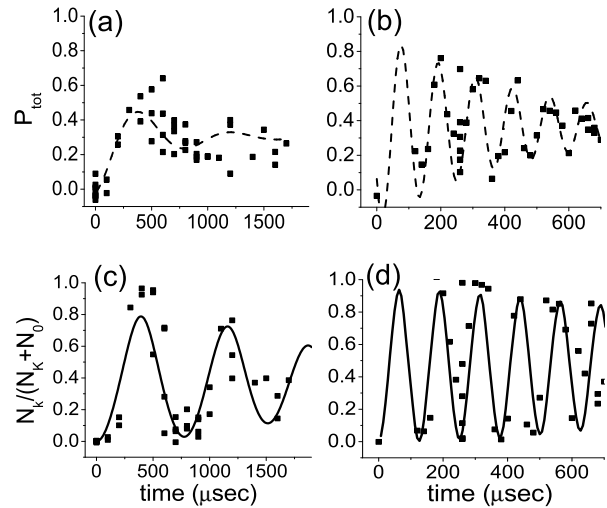}
\end{center}
\caption{Experimental oscillations between two coupled BECs (solid
boxes) (a-b) average momentum $P_{tot}$ at 1.2 (a) and 8.6 (b) kHz
oscillations. The oscillations are damped due to collisions. The
dashed lines in (a) and (b) are exponentially decaying oscillation
fits to the points. (c-d) Post-selection of the uncollided
fraction $N_k/(N_0+N_k)$ for the 1.2 (c) and 8.6 (d) kHz
oscillations. The solid lines in (c) and (d) are GPE simulations
with no fitting parameters.} \label{fig4}
\end{figure}

There are several points of interest in this figure. Firstly, we
note that the suppression of inhomogeneous effects is again
confirmed both in simulation and in experiment. This can be
observed by the minima of the simulation curve at $~3$ kHz driving
frequency. The experimental data agrees with this trend. Where the
observed decay rates of $P_{tot}$ and $N_k/(N_0+N_k)$ are 2 and 10
times smaller, respectively, than the calculated inhomogeneous
dephasing rate (as shown in Fig. 1).

Secondly, the GPE simulation dephasing rate ($\sim 0.1$ kHz) near
$3$ kHz driving frequency is significantly slower than the Doppler
dephasing rate ($0.6$ kHz). The experimental points agrees with
this trend. This suppression can be understood by considering the
system in the frame of reference of the travelling light potential
\cite{Stringari-lattice}. In this reference frame the condensate
is on the edge of the Brillion zone, and is repeatedly reflected
by the potential. Due to the strong lattice potential, there is a
broad region in momentum space for which the group velocity is
zero. Consequently, the two wavepackets do not separate,
suppressing the Doppler broadening as well. The reason for the
increase in the dephasing rate at higher driving frequencies
beyond $3$ kHz appears to result from the excitation of other
momentum modes along with wavepacket spreading in momentum space.

Thirdly, we note that we can use the measured decay rates of
$P_{tot}(t)$ and $N_k/(N_0+N_k)(t)$ to isolate and estimate the
decay due to collisions. We thus measure the collisional cross
section to be $7.1(\pm 1.8)\times 10^{-16} m^{2}$ in agreement
with the known \cite{collision-value} value of $8.37 \times
10^{-16} m^{2}$ \cite{collision-comment}.

\begin{figure}[h]
\begin{center}
\includegraphics[width=8cm]{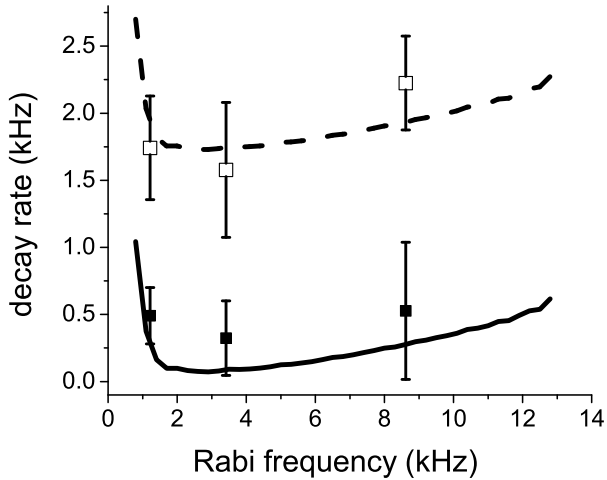}
\end{center}
\caption{Measured decay rates as a function of oscillation
frequency. The solid boxes are the exponential fit decay rates of
the oscillations in the post-selected $N_k/(N_0+N_k)$. The solid
line is the rate of decay fitted to the results of GPE
simulations. The open boxes are the fitted decay rates of average
momentum $P_{tot}$. The dashed line is the sum of the expected LDA
collision rate (1.66 kHz) with the decay rates from GPE
simulations (solid line).} \label{fig5}
\end{figure}

In conclusion, we measure a suppression of the mean-field shift
for a strong excitation of the BEC. We also observe an order of
magnitude suppression of the inhomogeneous broadening and Doppler
broadening mechanisms between strongly driven colliding
Bose-Einstein condensates. Furthermore, we measure a collisional
decoherence rate in agreement with that expected from previous
measurements.

In the future we hope to observe a splitting in a Bragg probe
spectrum from such a strongly driven system. We also hope to
measure a shift in the energy of the excitations due to
off-resonance Bragg pulses, in analogy with the ac Stark shift.
This shift may even modify the collision rate by shifting the
resonance energy sufficiently to influence the interaction with
the finite width collisional quasi-continuum \cite{kurizki}.

This work was supported in part by the Israel Ministry of Science,
the Israel Science Foundation and by Minerva foundation.


\begin{references}
\bibitem{lineshapes} For concreteness we
denote the dephasing time by the $1/e$ decay time of the cosine
transform of the lineshape. The dephasing rate is the inverse of
the dephasing time. This corresponds to an analytical estimate of
the decay rate of the Ramsey fringes.
\bibitem{collision-comment} For weak excitations,
at our experimental parameters, we expect a significant
suppression of the collisional cross section as compared to the
free particle value \cite{ours-beliaev} ($\sim 35\%$). However,
for strong excitations we do not expect such a large suppression,
since the perturbative assumptions do not apply.
\end{references}
\end{document}